\documentclass[fleqn,usenatbib,useAMS]{mnras}

\usepackage{graphicx}	% Including figure files
\usepackage{amsmath}	% Advanced maths commands
\usepackage{amssymb}	% Extra maths symbols
\usepackage{multicol}        % Multi-column entries in tables
\usepackage{bm}		% Bold maths symbols, including upright Greek
\usepackage{pdflscape}	% Landscape pages
\usepackage{csvsimple}

\usepackage[T1]{fontenc}
\usepackage{ae,aecompl}

\usepackage{newtxtext,newtxmath}
\title[The MeerKAT Thousand-Pulsar-Array programme]
{The Thousand-Pulsar-Array programme on MeerKAT I: Science objectives and first results}
\author[Johnston et al.]
{Simon Johnston$^{1}$\thanks{E-mail: simon.johnston@csiro.au},
A. Karastergiou$^{2,3,4}$, 
M. J. Keith$^{5}$,
X. Song$^{5}$,
P. Weltevrede$^{5}$, \newauthor
F. Abbate$^{6,7}$,
M. Bailes$^{8,9}$, 
S. Buchner$^{10}$,
F. Camilo$^{10}$,
M. Geyer$^{10}$,
B. Hugo$^{10,4}$,
A. Jameson$^{8,9}$, \newauthor
M. Kramer$^{11,5}$,
A. Parthasarathy$^{8,9}$,
D. J. Reardon$^{8,9}$,
A. Ridolfi$^{7,11}$,
M. Serylak$^{10,3}$, \newauthor
R. M. Shannon$^{8,9}$,
R. Spiewak$^{8,9}$,
W. van Straten$^{12}$,
V. Venkatraman Krishnan$^{11}$,
F. Jankowski$^{5}$, \newauthor
B. W. Meyers$^{13,14}$,
L. Oswald$^{2}$,
B. Posselt$^{2,15}$,
C. Sobey$^{16}$,
A. Szary$^{17,18}$,
J. van Leeuwen$^{17,19}$
\\
$^{1}$CSIRO Astronomy and Space Science, Australia Telescope National Facility, PO~Box~76, Epping NSW~1710, Australia\\
$^{2}$Oxford Astrophysics, Denys Wilkinson Building, Keble Road, Oxford, OX1 3RH, UK.\\
$^{3}$Physics Department, University of the Western Cape, Cape Town 7535, South Africa\\
$^{4}$Department of Physics and Electronics, Rhodes University, PO Box 94, Grahamstown 6140, South Africa\\
$^{5}$ Jodrell Bank Centre for Astrophysics, The University of Manchester, Alan Turing Building, Manchester, M13 9PL, United Kingdom\\
$^{6}$Dipartimento di Fisica `G. Occhialini', Universita` degli Studi Milano Bicocca, Piazza della Scienza 3, I-20126 Milano, Italy\\
$^{7}$INAF $-$ Osservatorio Astronomico di Cagliari, Via della Scienza 5, I-09047 Selargius (CA), Italy\\
$^8$Centre for Astrophysics and Supercomputing, Swinburne University of Technology, PO Box 218, Hawthorn, VIC 3122, Australia\\
$^9$ARC Centre of Excellence for Gravitational Wave Discovery (OzGrav)\\
$^{10}$South African Radio Astronomy Observatory (SARAO), 2 Fir Street, Black River Park, Observatory, Cape Town, 7925\\
$^{11}$Max-Planck-Institut fuer Radioastronomie, Auf dem Huegel 69, D-53121 Bonn, Germany.\\
$^{12}$Institute for Radio Astronomy \& Space Research, Auckland University of Technology, Private Bag 92006, Auckland 1142, New Zealand\\
$^{13}$Dept. of Physics and Astronomy, University of British Columbia, 6224 Agricultural Road, Vancouver, B.C., V6T 1Z1, Canada\\
$^{14}$International Centre for Radio Astronomy Research, Curtin University, Bentley, WA 6102, Australia\\
$^{15}$Department of Astronomy and Astrophysics,  The Pennsylvania State University, 525 Davey Laboratory, University Park, PA 16802, USA\\
$^{16}$CSIRO Astronomy and Space Science, PO Box 1130, Bentley, WA 6102, Australia\\
$^{17}$ASTRON, Netherlands Institute for Radio Astronomy, Oude Hoogeveensedijk 4, 7991 PD, Dwingeloo, The Netherlands\\
$^{18}$Janusz Gil Institute of Astronomy, University of Zielona G\'ora, Lubuska 2, 65-265 Zielona G\'ora, Poland\\
$^{19}$Anton Pannekoek Institute for Astronomy, University of Amsterdam, Science Park 904, 1098 XH Amsterdam, Netherlands
}
\date{Last updated; in original form}

\pubyear{2019}

\begin{document}
\label{firstpage}
\pagerange{\pageref{firstpage}--\pageref{lastpage}}
\maketitle

\begin{abstract}
We report here on initial results from the Thousand Pulsar Array (TPA) programme, part of the Large Survey Project ``MeerTime'' on the MeerKAT telescope. The interferometer is used in tied-array mode in the band from 856 to 1712~MHz, and the wide band coupled with the large collecting area and low receiver temperature make it an excellent telescope for the study of radio pulsars. The TPA is a 5 year project which aims to observe (a) more than 1000 pulsars to obtain high-fidelity pulse profiles, (b) some 500 of these pulsars over multiple epochs, (c) long sequences of single-pulse trains from several hundred pulsars. The scientific outcomes from the programme will include determination of pulsar geometries, the location of the radio emission within the pulsar magnetosphere, the connection between the magnetosphere and the crust and core of the star, tighter constraints on the nature of the radio emission itself as well as interstellar medium studies.
%The discovery of unknown phenomena given the high sensitivity of the telescope is a tantalising prospect.
First results presented here include updated dispersion measures, 26 pulsars with Faraday rotation measures derived for the first time and a description of interesting emission phenomena observed thus far.
\end{abstract}

\begin{keywords}
pulsars:general, instrumentation:interferometers
\end{keywords}

%%%%%%%%%%%%%%%%%%%%%%%%%%%%%%%%%%%%%%%%%%%%%%%%%%

%%%%%%%%%%%%%%%%% BODY OF PAPER %%%%%%%%%%%%%%%%%%
\section{Introduction}
Observations of radio pulsars have been a key plank of radio astronomy since their discovery in the late 1960s \citep{hbp+68}. Over the past few decades, pulsar observations have generally been the preserve of the world's largest single dishes, including the Parkes, Effelsberg, Arecibo, Lovell and Green Bank Telescopes. More recently, the power of interferometers has been harnessed towards pulsar observing including the Giant Metrewave Radio Telescope (GMRT) and the LOw Frequency ARray (LOFAR). The Parkes radio telescope has for decades been the most sensitive single dish in the southern hemisphere and has a long and glorious history in pulsar astronomy from the detection of the first pulsar glitch \citep{rm69}, through to the discovery of the double pulsar \citep{bdp+03,lbk+04}, Fast Radio Bursts \citep{lbm+07,tsb+13} and beyond. The commissioning of the MeerKAT telescope provides an alternative for observations of pulsars in the southern skies.

MeerKAT is an interferometer located in the Karoo, in South Africa's Northern Cape Province \citep{jon16}. In its current configuration it consists of 64 unblocked aperture dishes, each with an effective diameter of 13.97~m.
The 64 antennas in the array can be coherently phased together to form a highly sensitive `tied-array beam' which is used for pulsar observations. The L-band receiver has a band which extends from 856 to 1712~MHz, and has a receiver temperature of 18~K. The overall system equivalent flux density is 7.5~Jy, this compares to the Parkes value of 35~Jy in the same frequency band \citep{hmd+19}.

MeerTime is an approved Large Survey Project on the MeerKAT telescope for observing known radio pulsars \citep{bai19}. Pulsars can be used as a tool for cosmological tests of the stochastic gravitational wave background (e.g. \citealt{srl+15,abb+18}), tests of theories of gravity (e.g. \citealt{ksm+06,fkw12}), and understanding the equation of state of nuclear matter (e.g. \citealt{afw+13,of16}). In addition observations of pulsars give insight into the internal structure of neutron stars (e.g. \citealt{aeka18}), the workings of the pulsar magnetosphere (e.g. \citealt{dr15,ijw19}), and the structure of the Galaxy (e.g. \citealt{njkk08,ymw17}).

In the sections below we first outline the science case for the Thousand-Pulsar-Array programme within the MeerTime project. We then describe the initial observations and system configuration. We demonstrate that the wide band of the MeerKAT receivers are ideal for measurements of dispersion measure and rotation measure and show examples of the high signal-to-noise ratio  observations on individual pulsars. We end by briefly discussing the future of the project.

\section{The Thousand-Pulsar-Array programme}
The Thousand-Pulsar-Array (TPA) programme is part of MeerTime, the main project for observations of known pulsars on the MeerKAT telescope \citep{bai19}. The selection criteria for the TPA is that the pulsars (a) are south of declination +20\degr\, (b) have positional errors less than 2\arcsec\, and (c) are not recycled or millisecond pulsars. There are three main observational aims of the TPA. The first is to obtain a homogeneous set of polarimetric profiles for more than 1000 pulsars with sufficient signal-to-noise ratio to measure the standard parameters such as dispersion measure (DM) and rotation measure (RM), pulse widths and polarization properties. The second is to obtain observations with a regular cadence over a total time window of 5 years of at least 500 pulsars with sufficient observing time per observation to enable detection of pulse profile changes at the 10\% level. The third is to obtain sequences of single pulse trains in full polarization for more than 500 pulsars with sufficient signal-to-noise ratio to allow standard single-pulse analysis. Finally, the TPA schedule needs to be flexible to respond to glitches, state changes and other events which require follow-up observations. We estimate the total observing time to be $\sim$750~hr spread over 5 years, with the bulk of the time allocated to the regular monitoring project. 

The scientific goals that motivate these observations include the determination of the geometry for each pulsar, the location of the radio emission within the pulsar magnetosphere, the connection between the magnetosphere and the crust/core of the star, and the nature of the radio emission itself. In addition much can be gleaned of the properties of the interstellar medium (ISM) from the data. Each of these topics has a body of literature behind it; here we briefly summarise some open questions and a selection of recent references.

Fundamental to understanding individual pulsars, their properties and the pulsar population as a whole is the geometry of these objects and how it evolves with time. Determining the geometry of a large sample of pulsars remains fraught with uncertainties \citep{rwj15a} largely due to magnetospheric and relativistic distortions to the underlying simplicity of the rotating vector model \citep{rc69}. The evolution with time is a key theoretical question \citep{abp17} with somewhat conflicting observational results \citep{lgw+13,jk17}.

Radio emission arises within the pulsar magnetosphere, but the question regarding emission heights and the general 3D structure of the pulsar radio beam remains open (e.g. \citealt{psc15}). Conal structures \citep{ran93,omr19}, patchy structures \citep{lm88} or a hybrid \citep{kj07} have been considered. More recently, interesting ideas pertaining to fan beams have emerged from theory and observations \citep{dr15,dkl+19,okj19}.

The dynamic nature of the pulsar magnetosphere and the keys to the radio emission process are best revealed through the properties of individual pulses \citep{omr19,dyks19}. Observations of the drifting sub-pulse phenomenon \citep{bmm+16} and the associated carousel model \citep{sbw+19} along with interpretation of the bi-drifting phenomenon \citep{svl17,ww17} provide insights into the emission mechanism. Currently the southern hemisphere pulsars have not been well searched or studied for sub-pulse drifting and other phenomena.

\begin{table}
\caption{Table showing DM measurements from MeerKAT observations
that differ by more than 3~cm$^{-3}$pc from the catalogue value.}
\label{table:dm}
\begin{tabular}{lrrrr}
Pulsar & Cat. DM & uncert. & Measured DM & uncert.\\
& (cm$^{-3}$pc) & (cm$^{-3}$pc)  & (cm$^{-3}$pc) & (cm$^{-3}$pc) \\ \hline
J0804$-$3647   &  196      &      5    &  186.8    &  0.2  \\
J0818$-$3049   &  133.7    &    0.2    &  118.2    & 0.5  \\
J0902$-$6325   &   76      &      7    &   72.72   &  0.06 \\
J0905$-$4536   &   179.7   &           &   196.2   &  0.3  \\
J0932$-$5327   &   122     &     11    &  118.56   &  0.13 \\
J0952$-$3839   &   167     &      3    &   162.88  &  0.11 \\
J1020$-$6026   &   441.5   &     0.4   &   446.0   &  0.5  \\
J1231$-$4609   &   76      &      7    &   67.4    &  0.3  \\
J1316$-$6232   &   983.3   &     0.5   &   972.0   &  0.6  \\
J1427$-$4158   &    71     &      3    &   65.4    &  0.2 \\
J1504$-$5621   &   143     &      5    &   148.6   &  0.1  \\
J1525$-$5605   &   338     &      3    &   333.5   &  0.4  \\
J1527$-$5552   &   362.7   &      0.8  &   370.06  &  0.05 \\
J1536$-$3602   &   96      &      6    &   85.7    &  0.4  \\
J1604$-$4718   &   52.0    &     1.6   &   55.10   &  0.08 \\
J1651$-$7642   &   80      &     10    &   59.9    &  0.9  \\
J1739+0612     &   101.5   &     1.3   &   95.52   &  0.04 \\
J1759$-$2549   &   431     &      5    &   424.0   &  0.2  \\
J1811$-$2439   &   172.0   &    0.5    &   166.6   &  0.2  \\
J1814$-$1744   &   792     &     1.6   &   820     &  2    \\
J1827$-$0750   &   381     &     9     &   375.45  & 0.07  \\
J1832$-$0644   &   578     &      7    &   574.3   & 0.3   \\
J1839$-$0402   &   242     &      3    &   236.6   &  0.2  \\
J1839$-$0459   &   243     &      3    &   237.0   &  0.3  \\
J1840$-$1207   &   302.3   &     1.5   &   290.9   &  0.2  \\
J1842$-$0153   &   434     &      5    &  427.20   & 0.06  \\
J1842$-$0415   &   188     &      4    &  183.78   & 0.12  \\
J1844$-$0244   &   429     &      3    &  425.3    &  0.2  \\
J1848$-$0023   &   30.6    &      1    &   34.90   & 0.13  \\
J1849$-$0317   &   42.9    &     2.8   &   39.55   & 0.06  \\
J1850$-$0006   &   570     &    20     &   646     &  3    \\
\end{tabular}
\end{table}

Evidence has emerged that there is a complex connection between the magnetosphere and the crust and core of a neutron star \citep{pdh+18}. The ubiquity of, and link between, timing noise \citep{sc10,psj+19} and glitches \citep{fer+17,hkaa18} poses many open questions. In addition, mode-switching involves changes in the profile and the spin-down torque of some pulsars \citep{klo+06,lhk+10,bkj+16}, the reasons for which remain unclear \citep{khjs16,slk+19}. Understanding the frequency dependence of pulsar emission, and monitoring the correlations between magnetospheric properties and rotation are key to the design of the TPA survey.

The frequency- and time-dependent ISM response function that filters the radio emission from pulsars poses several open questions. These can be broadly categorized into two groups, those pertaining to the characteristics of the ISM itself \citep{wks+08,gkk+17,kcw+18}, and those that relate to how best to remove these effects from pulsar data to understand intrinsic pulsar properties \citep{wdv13,pen19}. The wide bandwidth of the TPA observations provides a large lever arm for measuring these effects.

\begin{table}
\caption{Rotation measures for pulsars without previous measurements.}
\label{table:rmnew}
\begin{tabular}{lrrr}
Pulsar & RM & uncertainty & $\langle B_{||} \rangle$\\
& (rad~m$^{-2}$) & (rad~m$^{-2}$) & ($\upmu$G) \\ \hline
J0630$-$0046 & 44 & 2 & 0.55 \\
J1104$-$6103 & $-$8.0 & 0.5 & $-$0.12 \\
J1130$-$6807 & $-$150 & 6 & $-$1.24 \\
J1308$-$5844 & 9.1 & 1.0 & 0.05 \\
J1434$-$6006 & $-$361 & 3 & $-$1.34 \\
J1457$-$5902 & 13.7 & 1.0 & 0.03 \\
J1502$-$5653 & $-$253.7 & 0.7 & $-$1.61\\
J1512$-$5431 & 5.3 & 0.6 & 0.03 \\
J1514$-$5925 & 6.4 & 1.0 & 0.04 \\
J1537$-$4912 & 0.0 & 3 & 0.00 \\
J1604$-$4718 & 1.6 & 1.1 & 0.03 \\
J1609$-$4616 & $-$115.3 & 1.4 & 0.94 \\
J1634$-$5640 & 159 & 5 & 1.32 \\
J1656$-$3621 & $-$23.4 & 0.5 & $-$0.12\\
J1716$-$4711 & $-$221.8 & 1.2 & $-$0.96 \\
J1722$-$4400 & $-$302.8 & 0.8 & $-$1.70 \\
J1726$-$4006 & 284 & 4 & 1.26 \\
J1732$-$4156 & $-$78 & 4 & $-$0.42 \\
J1758$-$1931 & 295 & 3 & 1.76 \\
J1808$-$1020 & 125.4 & 0.5 & 0.68 \\
J1808$-$1517 & $-$169 & 4 & $-$1.01\\
J1809$-$0743 & $-$44 & 2 & $-$0.23 \\
J1811$-$2439 & $-$82 & 2 & $-$0.59 \\
J1819$-$1008 & 124 & 2 & 0.38 \\
J1820$-$0509 & 77.6 & 0.8 & 0.92 \\
J1840$-$1207 & $-$126.2 & 1.4 & $-$0.52 \\
\end{tabular}
\end{table}

Combining TPA and multiwavelength observations enables further understanding of pulsar physics. High-energy observations can, for example, provide independent constraints on the pulsar geometry, either through the analysis of their multiwavelength pulse profiles \citep{Pierbattista2015,Giraud2019} or pulsar wind nebulae \citep{Ng2004,Klingler2018,Barkov2019}. Multiwavelength monitoring of unusual pulsars such as the radio-loud magnetars and transition objects can shed light onto potential magnetospheric differences between the diverse neutron star populations \citep{Camilo2018,djw+18}. Finally, the radio-X-ray correlation in mode-switching pulsars can put models of the pulsar magnetosphere to the test \citep{Hermsen2013,Rigoselli2019}.

The TPA is synergistic with related pulsar research being carried out world-wide. At the Parkes telescope, a sample of some 250 pulsars have been observed at 1.4~GHz monthly since 2007 \citep{wjm+10,jk18}. At the Lovell telescope, up to 40~yr of timing has been done on a very large sample of pulsars with regular cadence \citep{elsk11}. The UTMOST telescope, used as transit instrument since mid-2017, continues to monitor more than 300 pulsars at an observing frequency near 840~MHz \citep{jbv+19}. Finally the CHIME telescope observes every pulsar in the northern sky at transit every day. Each of these programmes has its own strengths and weaknesses and will provide complementarity to the TPA programme.

\section{Observations}
Observations for this programme commenced in 2019 February and as of the end of 2019 September a total of 315 pulsars have been observed. Generally the observations were carried out with at least 58 antennas used to form the tied-array beam. For the most part, data were recorded with 928 frequency channels between 896 and 1671~MHz with 1024 time bins across the pulsar period for each of the Stokes parameters. The data are coherently de-dispersed at the known DM using {\sc dspsr} \citep{vb11}, folded at the topocentric spin-period of the pulsar using its ephemeris\footnote{Ephemerides were obtained from the pulsar catalogue, available at https://www.atnf.csiro.au/research/pulsar/psrcat/} and written to disk every 8~seconds for the duration of the observation. In addition to the fold-mode data stream, a second stream is recorded at a sampling rate of 32~$\upmu$s for each channel, necessary for the study of single pulses. The data are written to disk in {\sc psrfits} format for data reduction with  the {\sc psrchive} software \citep{hvm04}. During the phase-up of the array, data are recorded to measure the relative gain and phase between the vertical and horizontal probes of the receiver per frequency channel. This allows for subsequent polarization calibration to be carried out. Observations of bright quasars are conducted to allow for flux calibration. In the near future these corrections are expected to be applied prior to writing the data to disk. A complete end-to-end description of the observing system is given in \citet{bai19}.

In order to achieve the goals of the TPA, careful consideration needs to be given to the total integration time per source per observing session. The integration time per source does not depend (only) on the desired signal-to-noise ratio. Rather, it is also important to gather enough rotations of the pulsar so that pulse-to-pulse fluctuations do not dominate the final profile. There is therefore a trade-off between raw sensitivity and observing time and this trade-off is different for the different parts of the project. For the `single-pulse census' it was decided to observe a minimum of 1024 pulses for all pulsars with a flux density at 1.4~GHz greater than 0.6~mJy. For the `polarimetric census', the integration time is determined via the pulse width, the flux density and the rotation period to ensure both high signal-to-noise ratio and minimum jitter noise. Finally, for the `repeat monitoring' campaign we require that the telescope can be configured into 2 equal-sized subarrays of 32 antennas. We observe of order 500 pulsars split between the two subarrays with the observing time set to minimise the jitter noise so that profile changes between epochs can be detected at the 10\% level per phase bin. A full description of these trade-offs will be presented in an upcoming publication.

\section{Initial Results}
\begin{figure}
\begin{tabular}{c}
\includegraphics[width=8cm]{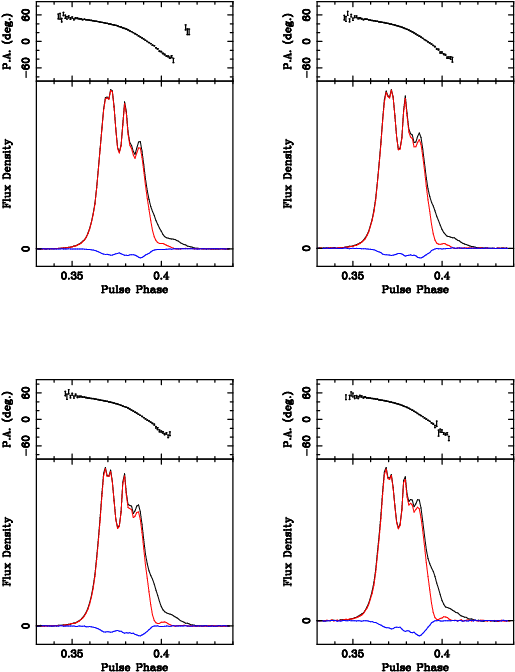} \\
\end{tabular}
\caption{Polarization profiles at centre frequencies of 993 (top left), 1188 (top right), 1381 (bottom left) and 1576 (bottom right) MHz for PSR~J0742$-$2822. Black denotes total intensity, red linear polarization and blue circular polarization. The top panel shows the position angle of the linear polarization.}
\label{J0742}
\end{figure}
\begin{figure}
\begin{tabular}{c}
\includegraphics[width=8cm]{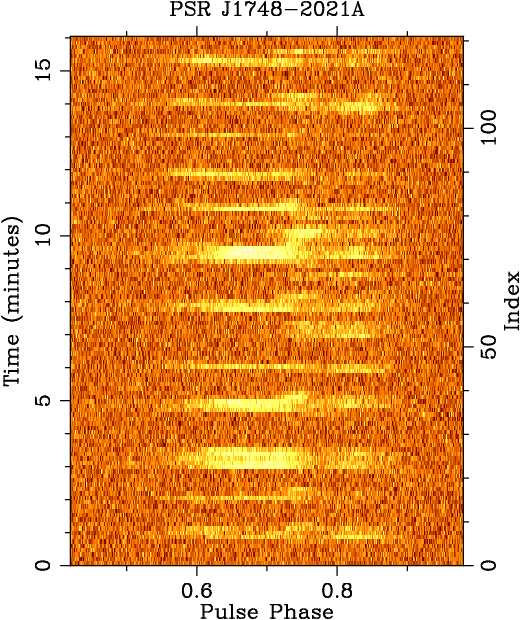}\\
\end{tabular}
\caption{Phase versus time showing nulling in the globular cluster pulsar
J1748$-$2021A. Each horizontal row comprises 8 seconds of data.}
\label{J1748}
\end{figure}
\subsection{Determination of Dispersion Measures}
We can exploit the wide bandwidth of the MeerKAT receiver coupled with the high signal-to-noise ratio of the observations to measure the dispersion measure to high accuracy. The DM is a measure of the total electron number density, $n_e$, along the line of sight, $L$, viz
\begin{equation}
{\rm DM} = \int^L_0 n_e(L)\,\, dL
\end{equation}
and has the units cm$^{-3}$pc.

Because of the dispersive effects of the interstellar medium, the high frequency radiation arrives at the telescope earlier than the low frequency radiation, and the DM can be derived by a measurement of this time delay viz
\begin{equation}
{\rm DM} = \frac{t_2-t_1}{4.149\,\,\,(\nu_1^{-2} - \nu_2^{-2})}
\end{equation}
with the observing frequencies, $\nu_1$ and $\nu_2$ in GHz and the arrival times $t_1$ and $t_2$ in ms. We therefore determine the DM for each of our pulsars by measuring the delay across the band. In practice this involves the use of the software package {\sc tempo2} \citep{hem06}. We first reduce the number of frequency channels by a factor of 32, and use a single noise-free template to produce a time-of-arrival (ToA) for each channel. {\sc tempo2} then fits a quadratic function to the ToAs to determine the DM and its uncertainty. Results are given in Table~\ref{table:dm} for those pulsars for which the DM differs by more than 3~cm$^{-3}$pc from the previously published value. We are aware that measurements of DM carried out in this standard fashion include a deviation from the true interstellar DM due to profile evolution with frequency and the use of a single template across the band. While previous works \citep[e.g.][]{pen19} have introduced techniques to compute frequency-resolved templates for high-precision pulsar timing, these templates only address the frequency dependence of the pulse shape. Frequency-dependent delays are due to a combination of interstellar dispersion and the intrinsic frequency-dependent direction of emission, which is a key aspect to the pulsar emission mechanism and remains an open problem that future TPA papers will address. 

We note the curious case of PSR~J0905$-$4536. In the original discovery paper \citep{mld+96} the DM was listed as $189\pm9$~cm$^{-3}$pc, yet in the timing paper of \citet{dsb+98} it became $116.8\pm0.2$~cm$^{-3}$pc which is clearly in error. \citet{bbb+12} spotted this error and computed the DM to be $179.7$~cm$^{-3}$pc (no uncertainty was given). Finally, our observation yields $196.2\pm0.2$~cm$^{-3}$pc, consistent (though with much smaller uncertainties) with the original detection! PSRs~J1020$-$6026, J1316$-$6232, J1527$-$5552, J1739+0612, J1814$-$1744 and J1840$-$1207 have DMs more than 5$\sigma$ away from their published values. We deem it unlikely that the pulsars' motion through the interstellar medium have caused these changes (see e.g. \citealt{pkj+13}). Most likely, the uncertainties were underestimated in the original publications due to the narrow bandwidths employed.

We have observed the bright pulsar PSR~J1057$-$5226 on 6 separate occasions between 2019 February and September. The DM uncertainty for a single 3 minute observation is only 0.003~cm$^{-3}$pc, about a factor of 10 better than the equivalent result using the Parkes telescope \citep{pkj+13}. There is a 2-$\sigma$ significant slope in the measured DMs  over the 200 day time span. The small uncertainties imply that multiple observations of bright pulsars over the 5 years of the thousand pulsar array programme should yield detections of DM variations in these slow pulsars, something which has hitherto only been possible for a handful of objects. It may also be possible to exploit the wide bandwidth to measure the `chromatic' DM effects pointed out by \citet{css16}.

\begin{figure}
\begin{tabular}{c}
\includegraphics[width=0.95\hsize]{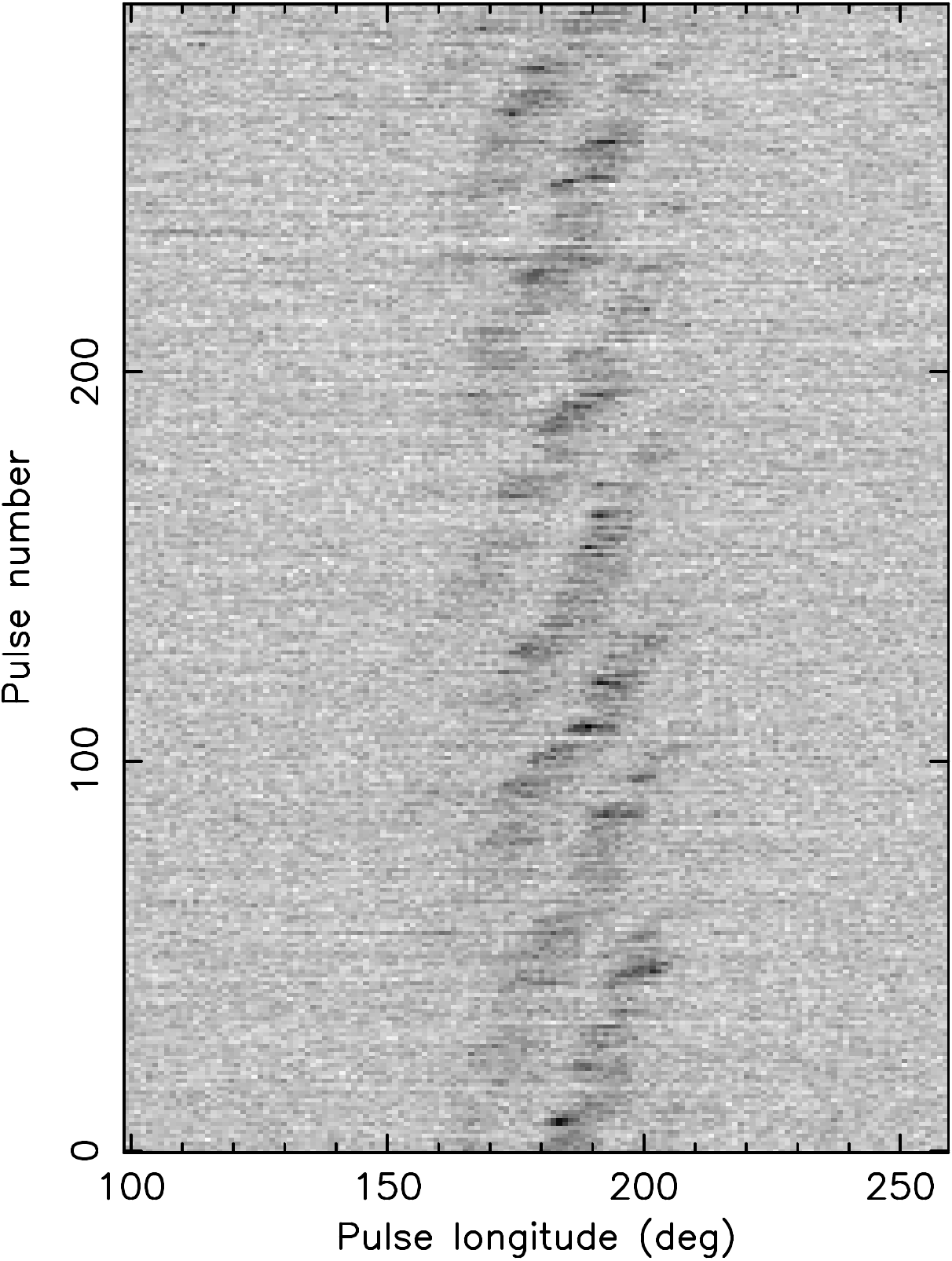} \\
\end{tabular}
\caption{The drifting subpulses of PSR J1750$-$3503 are revealed in the pulse stack of 295 successive pulses.}
\label{J1750}
\end{figure}
\begin{figure}
\begin{tabular}{c}
\includegraphics[width=8cm]{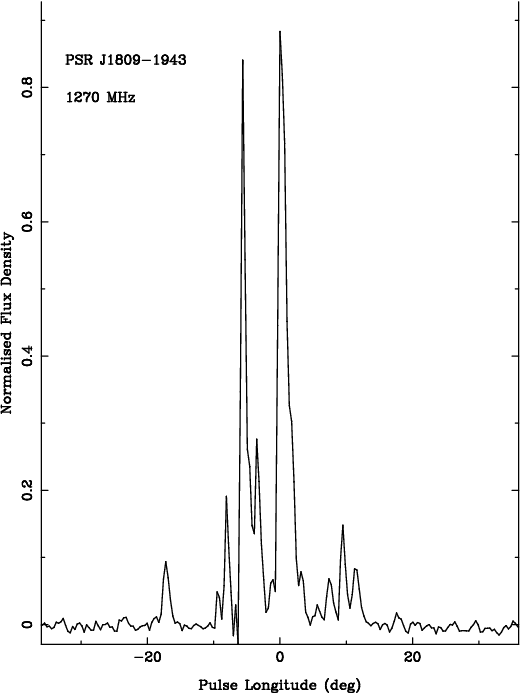} \\
\end{tabular}
\caption{A Stokes I profile of a single pulse of the magnetar XTE~J1810$-$197
(PSR~J1809$-$1943) taken on 2019 March 15.}
\label{magnetar}
\end{figure}
\subsection{Determination of Rotation Measures}
Pulsars are often highly linearly polarized (e.g. \citealt{jk18}). As the radiation traverses the interstellar medium it suffers Faraday rotation which rotates the position angle of the linearly polarized emission. The rotation measure (RM) is the product of the magnetic field parallel to the  line of sight, $\langle B_{||} \rangle$ and the electron density integrated over the path length to the pulsar:
\begin{equation}
{\rm RM} = 0.81 \int^L_0 n_e(L)\,\, B_{||}(L)\,\, dL
\end{equation}
With $L$ in pc, $n_e$ in cm$^{-3}$ and $B_{||}$ in $\upmu$G then the units of RM are rad~m$^{-2}$. The combination of the RM and the DM yields the average magnetic field strength parallel to the line of sight.

The RM can be determined by measuring the shift in position angle, $\Delta\theta$, at different observing wavelengths, $\lambda_1$ and $\lambda_2$:
\begin{equation}
{\rm RM} = \frac{\Delta\theta}{\lambda_1^2 - \lambda_1^2}
\end{equation}
In practice we use a combination of methods to compute the RM from the data. One such method is to use trial RMs to de-Faraday the data and then measure the flux density of the linear polarization; the RM at which this peaks is the best RM. This method is similar to the so-called RM synthesis methods used by e.g. \citet{sbg+19}. We also used the method outlined in \citet{njkk08} by measuring the slope to a straight line fit of $\Delta\theta$ versus $\lambda^2$.  Finally once the RM is known approximately the error bar can be reduced by reducing the number of channels to two and measuring the shift in PA between the two halves of the band. This is the method used by \citet{hmvd19}.

\begin{figure}
\begin{tabular}{c}
\includegraphics[width=8cm]{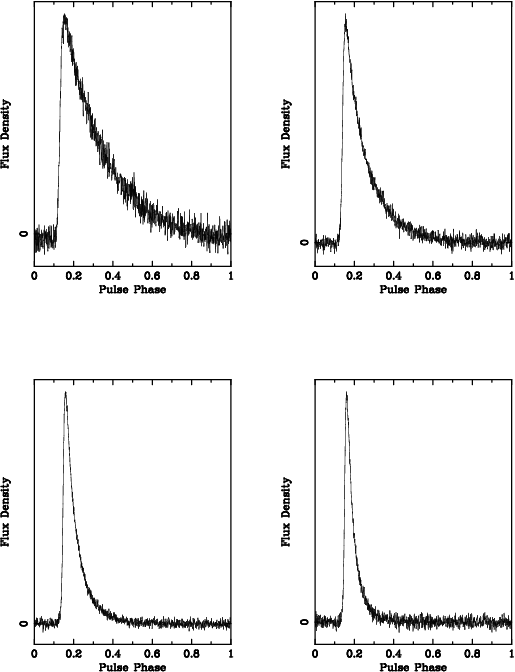} \\
\end{tabular}
\caption{Profiles of PSR~J1818$-$1422 at centre frequencies of 987 (top left), 1183 (top right), 1390 (bottom left) and 1600 (bottom right) MHz showing the strong dependency of interstellar scattering with frequency.}
\label{J1818}
\end{figure}
The RM values for 26 pulsars without previous measurements are given in Table~\ref{table:rmnew} along with the derived $\langle B_{||} \rangle$ using the pulsar's DM. Uncertainties are $1-\sigma$. Where necessary we also derived an improved DM, as incorrect DM values can result in incorrect RM measurements (e.g. \citealt{ijw19}). These RM values have been corrected for the contribution
from the Earth's ionosphere. This contribution was estimated for each pulsar using the publicly-available code ionFR\footnote{https://github.com/csobey/ionFR/} (see \citealt{ssh+13}), with inputs from the International Geomagnetic Reference Field\footnote{https://www.ngdc.noaa.gov/IAGA/vmod/igrf.html}
%(IGRF version 12; Thébault et al. 2015
and International GNSS Service vertical total electron content maps\footnote{ftp://cddis.nasa.gov/pub/gps/products/ionex/}.
%(e.g. Hernández-Pajares et al. 2009)
In all except three cases, the ionospheric contribution to the observed RM is between $-$0.2 and $-$0.75 rad~m$^{-2}$. For the observations of PSRs~J0630$-$0046, J1104$-$6103 and J1514$-$4834 the ionosphere contributes $-1.3\pm0.2$, $-2.0\pm0.1$, and $-1.8\pm0.1$ rad~m$^{-2}$, respectively.

\subsection{Pulsars of interest}
{\bf PSR~J0045$-$7319:} This pulsar is in the Small Magellanic Cloud and was found to be in orbit about a main sequence B-star \citep{kjb+94,bbs+95}. With a low flux density at 1400~MHz, timing at the Parkes telescope has been difficult and observations of the pulsar ceased a decade ago. We carried out two 8 minute observations with the MeerKAT array in 2019 February and May. The pulsar was easily detected with a signal-to-noise ratio of 25, and is clearly much brighter at the lower frequencies. This promises well for future prospects of determining the spin-orbit coupling in the system \citep{kbm+96}.

\noindent
{\bf PSR~J0540$-$6919:} This pulsar, with a spin period of only 50~ms, is located in the Large Magellanic Cloud and is known to emit giant pulses \citep{jr03}. Observations with the Parkes telescope detect a giant pulse for every 30~mins of observing time \citep{jrmz04}, however the sensitivity of MeerKAT means that we detect giant pulses with signal-to-noise ratio $>7$ from the pulsar at a rate greater than 1 per minute. This has allowed us to determine better logN-logS statistics and for the first time we have measured the polarization properties of the giant pulses. Details will be given in an upcoming paper.

\begin{figure}
\begin{tabular}{c}
\includegraphics[width=8cm]{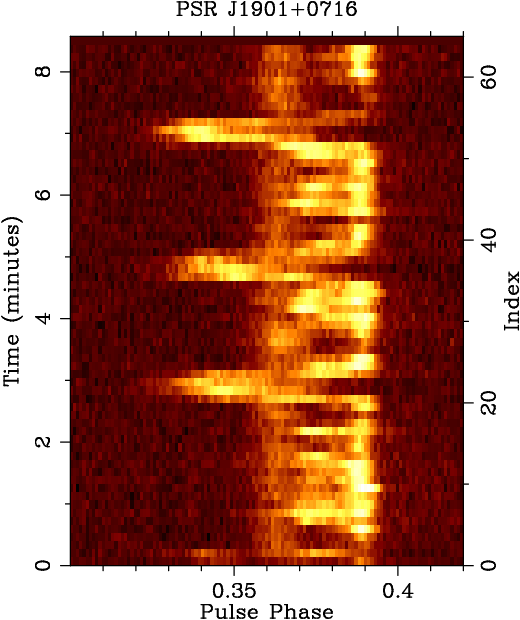} \\
\end{tabular}
\caption{Time versus phase for an 8 minute observation of
PSR~J1901+0716 showing several transitions between the two states. Each horizontal row represents 8~s of observing.}
\label{J1901}
\end{figure}
\noindent
{\bf PSR~J0742--2822:} Polarization calibration of the tied-array remains an on-going challenge at the present time. This pulsar makes a useful test case; it is bright, highly linearly polarized and has a known absolute position angle of the linear polarization \citep{jhv+05}. Figure~\ref{J0742} shows the polarization profile at 4 different frequencies across the total bandwidth of the receiver. The majority of the features in these profiles are correct, in particular the high fraction of linear polarization with depolarization on the trailing edge of the profile, the negative sign of the circular polarization and the swing of the position angle. Work is in progress to determine the instrumental leakage terms, and to apply the polarization corrections directly at the time of observation without the need to off-line calibration.

\noindent
{\bf PSR~J1748-2021A:}
This pulsar is in the globular cluster NGC~6440 yet has a long spin period of 288~ms. Studies by \citet{lmd96} and \citet{frb+08} did not mention the fact that this pulsar nulls. Our 15 minute observation of the pulsar taken on 2019 July 14 (see Figure~\ref{J1748}) shows evidence for nulling and mode switching between the two principal components in the pulse profile. Like all pulsars in the TPA programme with this type of behaviour, further investigation is warranted. 

\noindent
{\bf PSR~J1750$-$3503:} The pulse phase of the emission of this 684~ms pulsar changes in a highly periodic fashion from pulse to pulse (see Fig. \ref{J1750}), something not previously reported for this pulsar. The drifting sub-pulse pattern repeats itself every $\sim50$ rotations of the star, which is an exceptionally long timescale compared to other pulsars of this kind (e.g. \citealt{ran86,wes06,bmm+16}). This could well be related to to the fact that the characteristic age of this pulsar is very large, $\sim280$~Myr. Such a correlation could indeed be expected from a carousel model \citep{rs75}. This observation highlights the importance of obtaining a TPA legacy dataset with long enough sequences of single pulses for the brightest pulsars to quantify the pulse-to-pulse variability.

\noindent
{\bf XTE~J1810$-$197:} Following an X-ray outburst \citep{gha+19}, this magnetar recently switched back on in the radio and was observed with the Parkes telescope by \citet{dlb+19} and with the Lovell and Effelsberg telescopes by \citet{lld+19}. Figure~\ref{magnetar} shows a single pulse from the magnetar in data taken on 2019 March 15 showing the spiky nature of its emission.

\noindent
{\bf PSR~J1818$-$1422:} The wide band of the MeerKAT receiver is ideal for study of the scattering imposed by the interstellar medium as the scattering time will vary by a factor of $\sim$20 from the lowest frequency to the highest. Figure~\ref{J1818} shows the profile of PSR~J1818$-$1422 at 4 different frequencies across the observing band. Details of the functional form of the scattering and the frequency dependence for this and other pulsars from the sample will be presented elsewhere.

\noindent
{\bf PSR~J1901+0716:}  This pulsar is known to undergo episodic changes in its emission profile as first identified by \citet{rrw06} and studied in detail by \citet{worw16} and \citet{psw+16}. Figure~\ref{J1901} displays 8 minutes of data on the pulsar clearly showing the transition between the states. Long term monitoring of this pulsar will help our understanding of the magnetospheric variability seen at various time-scales, and the link between magnetospheric activity and the spin-down of the star.

\section{Summary}
MeerKAT is delivering high quality and high fidelity pulsar data. We have shown that the polarization properties of the instrument across its 1~GHz of bandwidth are excellent and comparable with single dishes (e.g. Figure~\ref{J0742}). The large bandwidth and high sensitivity allow us to measure RMs with high precision even for weak pulsars (e.g. PSR~J1722$-$4400 with a flux density of only 0.2~mJy, see Table~\ref{table:rmnew}). We have shown that previous DM measurements made from observations of weak pulsars over a narrow bandwidth can be improved with MeerKAT observations (see Table~\ref{table:dm}). Finally we showed an example of a nulling pulsar (Figure~\ref{J1748}) and a drifting pulsar (Figure~\ref{J1750}) not previously known. In the near future, the ability to sub-array the 64 antennas, and the installation of the low frequency band receivers (544 to 1088~MHz) will only enhance the capabilities of this programme. Data from the TPA will be made publicly available once quality control and calibration is complete.

\section*{Acknowledgements}
The MeerKAT telescope is operated by the South African Radio Astronomy Observatory, which is a facility of the National Research Foundation, an agency of the Department of Science and Innovation.
FA and AR acknowledge support through the research grant ``iPeska'' (P.I. Andrea Possenti) funded under the INAF national call Prin-SKA/CTA approved with the Presidential Decree 70/2016.
MB, RS and RMS acknowledge support through ARC grants FL150100148 and CE170100004.
AS and JvL acknowledge funding from the Netherlands Organisation for Scientific Research (NWO) under project ``CleanMachine'' (614.001.301), and from the ERC under the European Union's Seventh Framework Programme (FP/2007-2013) / ERC Grant Agreement n. 617199.
FJ acknowledges funding from the ERC under the European Union's Horizon 2020 research and innovation programme (grant agreement No. 694745).
LO acknowledges funding from the UK Science and Technology Facilities Council (STFC) Grant Code ST/R505006/1. AK and BP also acknowledge funding from the STFC consolidated grant to Oxford Astrophysics. 
Pulsar research at Jodrell Bank Centre for Astrophysics and Jodrell Bank
Observatory is also supported by a consolidated grant from STFC.
%%%%%%%%%%%%%%%%%%%%%%%%%%%%%%%%%%%%%%%%%%%%%%%%%%



\bibliographystyle{mnras}
\bibliography{tpa1} % if your bibtex file is called example.bib

\begin{thebibliography}{}
\makeatletter
\relax
\def\mn@urlcharsother{\let\do\@makeother \do\$\do\&\do\#\do\^\do\_\do\%\do\~}
\def\mn@doi{\begingroup\mn@urlcharsother \@ifnextchar [ {\mn@doi@}
  {\mn@doi@[]}}
\def\mn@doi@[#1]#2{\def\@tempa{#1}\ifx\@tempa\@empty \href
  {http://dx.doi.org/#2} {doi:#2}\else \href {http://dx.doi.org/#2} {#1}\fi
  \endgroup}
\def\mn@eprint#1#2{\mn@eprint@#1:#2::\@nil}
\def\mn@eprint@arXiv#1{\href {http://arxiv.org/abs/#1} {{\tt arXiv:#1}}}
\def\mn@eprint@dblp#1{\href {http://dblp.uni-trier.de/rec/bibtex/#1.xml}
  {dblp:#1}}
\def\mn@eprint@#1:#2:#3:#4\@nil{\def\@tempa {#1}\def\@tempb {#2}\def\@tempc
  {#3}\ifx \@tempc \@empty \let \@tempc \@tempb \let \@tempb \@tempa \fi \ifx
  \@tempb \@empty \def\@tempb {arXiv}\fi \@ifundefined
  {mn@eprint@\@tempb}{\@tempb:\@tempc}{\expandafter \expandafter \csname
  mn@eprint@\@tempb\endcsname \expandafter{\@tempc}}}

\bibitem[\protect\citeauthoryear{{Antoniadis} et~al.,}{{Antoniadis}
  et~al.}{2013}]{afw+13}
{Antoniadis} J.,  et~al., 2013, Science, 340, 448

\bibitem[\protect\citeauthoryear{{Antonopoulou}, {Espinoza}, {Kuiper}  \&
  {Andersson}}{{Antonopoulou} et~al.}{2018}]{aeka18}
{Antonopoulou} D.,  {Espinoza} C.~M.,  {Kuiper} L.,   {Andersson} N.,  2018,
  MNRAS, 473, 1644

\bibitem[\protect\citeauthoryear{{Arzamasskiy}, {Beskin}  \&
  {Pirov}}{{Arzamasskiy} et~al.}{2017}]{abp17}
{Arzamasskiy} L.~I.,  {Beskin} V.~S.,   {Pirov} K.~K.,  2017, MNRAS, 466, 2325

\bibitem[\protect\citeauthoryear{{Arzoumanian} et~al.,}{{Arzoumanian}
  et~al.}{2018}]{abb+18}
{Arzoumanian} Z.,  et~al., 2018, ApJ, 859, 47

\bibitem[\protect\citeauthoryear{{Bailes} et~al.,}{{Bailes}
  et~al.}{2019}]{bai19}
{Bailes} M.,  et~al., 2019, PASA. Submitted

\bibitem[\protect\citeauthoryear{{Barkov}, {Lyutikov}  \&
  {Khangulyan}}{{Barkov} et~al.}{2019}]{Barkov2019}
{Barkov} M.~V.,  {Lyutikov} M.,   {Khangulyan} D.,  2019, MNRAS, 484, 4760

\bibitem[\protect\citeauthoryear{{Basu}, {Mitra}, {Melikidze}, {Maciesiak},
  {Skrzypczak}  \& {Szary}}{{Basu} et~al.}{2016}]{bmm+16}
{Basu} R.,  {Mitra} D.,  {Melikidze} G.~I.,  {Maciesiak} K.,  {Skrzypczak} A.,
   {Szary} A.,  2016, ApJ, 833, 29

\bibitem[\protect\citeauthoryear{{Bates} et~al.,}{{Bates}
  et~al.}{2012}]{bbb+12}
{Bates} S.~D.,  et~al., 2012, MNRAS, 427, 1052

\bibitem[\protect\citeauthoryear{{Bell}, {Bessell}, {Stappers}, {Bailes}  \&
  {Kaspi}}{{Bell} et~al.}{1995}]{bbs+95}
{Bell} J.~F.,  {Bessell} M.~S.,  {Stappers} B.~W.,  {Bailes} M.,   {Kaspi}
  V.~M.,  1995, ApJ, 447, L117

\bibitem[\protect\citeauthoryear{{Brook}, {Karastergiou}, {Johnston}, {Kerr},
  {Shannon}  \& {Roberts}}{{Brook} et~al.}{2016}]{bkj+16}
{Brook} P.~R.,  {Karastergiou} A.,  {Johnston} S.,  {Kerr} M.,  {Shannon}
  R.~M.,   {Roberts} S.~J.,  2016, MNRAS, 456, 1374

\bibitem[\protect\citeauthoryear{{Burgay} et~al.,}{{Burgay}
  et~al.}{2003}]{bdp+03}
{Burgay} M.,  et~al., 2003, Nature, 426, 531

\bibitem[\protect\citeauthoryear{{Camilo} et~al.,}{{Camilo}
  et~al.}{2018}]{Camilo2018}
{Camilo} F.,  et~al., 2018, ApJ, 856, 180

\bibitem[\protect\citeauthoryear{{Cordes}, {Shannon}  \& {Stinebring}}{{Cordes}
  et~al.}{2016}]{css16}
{Cordes} J.~M.,  {Shannon} R.~M.,   {Stinebring} D.~R.,  2016, ApJ, 817, 16

\bibitem[\protect\citeauthoryear{{D'Amico}, {Stappers}, {Bailes}, {Martin},
  {Bell}, {Lyne}  \& {Manchester}}{{D'Amico} et~al.}{1998}]{dsb+98}
{D'Amico} N.,  {Stappers} B.~W.,  {Bailes} M.,  {Martin} C.~E.,  {Bell} J.~F.,
  {Lyne} A.~G.,   {Manchester} R.~N.,  1998, MNRAS, 297, 28

\bibitem[\protect\citeauthoryear{{Dai} et~al.,}{{Dai} et~al.}{2018}]{djw+18}
{Dai} S.,  et~al., 2018, MNRAS, 480, 3584

\bibitem[\protect\citeauthoryear{{Dai} et~al.,}{{Dai} et~al.}{2019}]{dlb+19}
{Dai} S.,  et~al., 2019, ApJ, 874, L14

\bibitem[\protect\citeauthoryear{{Desvignes} et~al.,}{{Desvignes}
  et~al.}{2019}]{dkl+19}
{Desvignes} G.,  et~al., 2019, Science, 365, 1013

\bibitem[\protect\citeauthoryear{{Dyks}}{{Dyks}}{2019}]{dyks19}
{Dyks} J.,  2019, MNRAS, 488, 2018

\bibitem[\protect\citeauthoryear{{Dyks} \& {Rudak}}{{Dyks} \&
  {Rudak}}{2015}]{dr15}
{Dyks} J.,  {Rudak} B.,  2015, MNRAS, 446, 2505

\bibitem[\protect\citeauthoryear{{Espinoza}, {Lyne}, {Stappers}  \&
  {Kramer}}{{Espinoza} et~al.}{2011}]{elsk11}
{Espinoza} C.~M.,  {Lyne} A.~G.,  {Stappers} B.~W.,   {Kramer} M.,  2011,
  MNRAS, 414, 1679

\bibitem[\protect\citeauthoryear{{Freire}, {Ransom}, {B{\'e}gin}, {Stairs},
  {Hessels}, {Frey}  \& {Camilo}}{{Freire} et~al.}{2008}]{frb+08}
{Freire} P.~C.~C.,  {Ransom} S.~M.,  {B{\'e}gin} S.,  {Stairs} I.~H.,
  {Hessels} J.~W.~T.,  {Frey} L.~H.,   {Camilo} F.,  2008, ApJ, 675, 670

\bibitem[\protect\citeauthoryear{{Freire}, {Kramer}  \& {Wex}}{{Freire}
  et~al.}{2012}]{fkw12}
{Freire} P.~C.~C.,  {Kramer} M.,   {Wex} N.,  2012, Classical and Quantum
  Gravity, 29, 184007

\bibitem[\protect\citeauthoryear{{Fuentes}, {Espinoza}, {Reisenegger}, {Shaw},
  {Stappers}  \& {Lyne}}{{Fuentes} et~al.}{2017}]{fer+17}
{Fuentes} J.~R.,  {Espinoza} C.~M.,  {Reisenegger} A.,  {Shaw} B.,  {Stappers}
  B.~W.,   {Lyne} A.~G.,  2017, A\&A, 608, A131

\bibitem[\protect\citeauthoryear{{Geyer} et~al.,}{{Geyer}
  et~al.}{2017}]{gkk+17}
{Geyer} M.,  et~al., 2017, MNRAS, 470, 2659

\bibitem[\protect\citeauthoryear{{Giraud} \& {P{\'e}tri}}{{Giraud} \&
  {P{\'e}tri}}{2019}]{Giraud2019}
{Giraud} Q.,  {P{\'e}tri} J.,  2019, MNRAS. Submitted., p. arXiv:1910.01555

\bibitem[\protect\citeauthoryear{{Gotthelf} et~al.,}{{Gotthelf}
  et~al.}{2019}]{gha+19}
{Gotthelf} E.~V.,  et~al., 2019, ApJ, 874, L25

\bibitem[\protect\citeauthoryear{{Han}, {Manchester}, {van Straten}  \&
  {Demorest}}{{Han} et~al.}{2018}]{hmvd19}
{Han} J.~L.,  {Manchester} R.~N.,  {van Straten} W.,   {Demorest} P.,  2018,
  ApJSS, 234, 11

\bibitem[\protect\citeauthoryear{{Haskell}, {Khomenko}, {Antonelli}  \&
  {Antonopoulou}}{{Haskell} et~al.}{2018}]{hkaa18}
{Haskell} B.,  {Khomenko} V.,  {Antonelli} M.,   {Antonopoulou} D.,  2018,
  MNRAS, 481, L146

\bibitem[\protect\citeauthoryear{{Hermsen} et~al.,}{{Hermsen}
  et~al.}{2013}]{Hermsen2013}
{Hermsen} W.,  et~al., 2013, Science, 339, 436

\bibitem[\protect\citeauthoryear{{Hewish}, {Bell}, {Pilkington}, {Scott}  \&
  {Collins}}{{Hewish} et~al.}{1968}]{hbp+68}
{Hewish} A.,  {Bell} S.~J.,  {Pilkington} J.~D.~H.,  {Scott} P.~F.,   {Collins}
  R.~A.,  1968, Nature, 217, 709

\bibitem[\protect\citeauthoryear{{Hobbs}, {Edwards}  \& {Manchester}}{{Hobbs}
  et~al.}{2006}]{hem06}
{Hobbs} G.~B.,  {Edwards} R.~T.,   {Manchester} R.~N.,  2006, MNRAS, 369, 655

\bibitem[\protect\citeauthoryear{{Hobbs} et~al.,}{{Hobbs}
  et~al.}{2019}]{hmd+19}
{Hobbs} G.,  et~al., 2019, PASA. Submitted, p. arXiv:1911.00656

\bibitem[\protect\citeauthoryear{{Hotan}, {van Straten}  \&
  {Manchester}}{{Hotan} et~al.}{2004}]{hvm04}
{Hotan} A.~W.,  {van Straten} W.,   {Manchester} R.~N.,  2004, PASA, 21, 302

\bibitem[\protect\citeauthoryear{{Ilie}, {Johnston}  \& {Weltevrede}}{{Ilie}
  et~al.}{2019}]{ijw19}
{Ilie} C.~D.,  {Johnston} S.,   {Weltevrede} P.,  2019, MNRAS, 483, 2778

\bibitem[\protect\citeauthoryear{{Jankowski} et~al.,}{{Jankowski}
  et~al.}{2019}]{jbv+19}
{Jankowski} F.,  et~al., 2019, MNRAS, 484, 3691

\bibitem[\protect\citeauthoryear{{Johnston} \& {Karastergiou}}{{Johnston} \&
  {Karastergiou}}{2017}]{jk17}
{Johnston} S.,  {Karastergiou} A.,  2017, MNRAS, 467, 3493

\bibitem[\protect\citeauthoryear{{Johnston} \& {Kerr}}{{Johnston} \&
  {Kerr}}{2018}]{jk18}
{Johnston} S.,  {Kerr} M.,  2018, MNRAS, 474, 4629

\bibitem[\protect\citeauthoryear{{Johnston} \& {Romani}}{{Johnston} \&
  {Romani}}{2003}]{jr03}
{Johnston} S.,  {Romani} R.~W.,  2003, ApJ, 590, L95

\bibitem[\protect\citeauthoryear{{Johnston}, {Romani}, {Marshall}  \&
  {Zhang}}{{Johnston} et~al.}{2004}]{jrmz04}
{Johnston} S.,  {Romani} R.~W.,  {Marshall} F.~E.,   {Zhang} W.,  2004, MNRAS,
  355, 31

\bibitem[\protect\citeauthoryear{{Johnston}, {Hobbs}, {Vigeland}, {Kramer},
  {Weisberg}  \& {Lyne}}{{Johnston} et~al.}{2005}]{jhv+05}
{Johnston} S.,  {Hobbs} G.,  {Vigeland} S.,  {Kramer} M.,  {Weisberg} J.~M.,
  {Lyne} A.~G.,  2005, MNRAS, 364, 1397

\bibitem[\protect\citeauthoryear{{Jonas} \& {MeerKAT Team}}{{Jonas} \& {MeerKAT
  Team}}{2016}]{jon16}
{Jonas} J.,  {MeerKAT Team} 2016, in Proceedings of MeerKAT Science: On the
  Pathway to the SKA. 25-27 May. p.~1

\bibitem[\protect\citeauthoryear{{Karastergiou} \& {Johnston}}{{Karastergiou}
  \& {Johnston}}{2007}]{kj07}
{Karastergiou} A.,  {Johnston} S.,  2007, MNRAS, 380, 1678

\bibitem[\protect\citeauthoryear{{Kaspi}, {Johnston}, {Bell}, {Manchester},
  {Bailes}, {Bessell}, {Lyne}  \& {D'Amico}}{{Kaspi} et~al.}{1994}]{kjb+94}
{Kaspi} V.~M.,  {Johnston} S.,  {Bell} J.~F.,  {Manchester} R.~N.,  {Bailes}
  M.,  {Bessell} M.,  {Lyne} A.~G.,   {D'Amico} N.,  1994, ApJ, 423, L43

\bibitem[\protect\citeauthoryear{{Kaspi}, {Bailes}, {Manchester}, {Stappers}
  \& {Bell}}{{Kaspi} et~al.}{1996}]{kbm+96}
{Kaspi} V.~M.,  {Bailes} M.,  {Manchester} R.~N.,  {Stappers} B.~W.,   {Bell}
  J.~F.,  1996, Nature, 381, 584

\bibitem[\protect\citeauthoryear{{Kerr}, {Hobbs}, {Johnston}  \&
  {Shannon}}{{Kerr} et~al.}{2016}]{khjs16}
{Kerr} M.,  {Hobbs} G.,  {Johnston} S.,   {Shannon} R.~M.,  2016, MNRAS, 455,
  1845

\bibitem[\protect\citeauthoryear{{Kerr}, {Coles}, {Ward}, {Johnston}, {Tuntsov}
   \& {Shannon}}{{Kerr} et~al.}{2018}]{kcw+18}
{Kerr} M.,  {Coles} W.~A.,  {Ward} C.~A.,  {Johnston} S.,  {Tuntsov} A.~V.,
  {Shannon} R.~M.,  2018, \mnras, 474, 4637

\bibitem[\protect\citeauthoryear{{Klingler}, {Kargaltsev}, {Pavlov}, {Ng},
  {Beniamini}  \& {Volkov}}{{Klingler} et~al.}{2018}]{Klingler2018}
{Klingler} N.,  {Kargaltsev} O.,  {Pavlov} G.~G.,  {Ng} C.~Y.,  {Beniamini} P.,
    {Volkov} I.,  2018, ApJ, 861, 5

\bibitem[\protect\citeauthoryear{{Kramer}, {Lyne}, {O'Brien}, {Jordan}  \&
  {Lorimer}}{{Kramer} et~al.}{2006a}]{klo+06}
{Kramer} M.,  {Lyne} A.~G.,  {O'Brien} J.~T.,  {Jordan} C.~A.,   {Lorimer}
  D.~R.,  2006a, Science, 312, 549

\bibitem[\protect\citeauthoryear{{Kramer} et~al.,}{{Kramer}
  et~al.}{2006b}]{ksm+06}
{Kramer} M.,  et~al., 2006b, Science, 314, 97

\bibitem[\protect\citeauthoryear{{Levin} et~al.,}{{Levin}
  et~al.}{2019}]{lld+19}
{Levin} L.,  et~al., 2019, MNRAS, 488, 5251

\bibitem[\protect\citeauthoryear{{Lorimer}, {Bailes}, {McLaughlin}, {Narkevic}
  \& {Crawford}}{{Lorimer} et~al.}{2007}]{lbm+07}
{Lorimer} D.~R.,  {Bailes} M.,  {McLaughlin} M.~A.,  {Narkevic} D.~J.,
  {Crawford} F.,  2007, Science, 318, 777

\bibitem[\protect\citeauthoryear{{Lyne} \& {Manchester}}{{Lyne} \&
  {Manchester}}{1988}]{lm88}
{Lyne} A.~G.,  {Manchester} R.~N.,  1988, MNRAS, 234, 477

\bibitem[\protect\citeauthoryear{{Lyne}, {Manchester}  \& {D'Amico}}{{Lyne}
  et~al.}{1996}]{lmd96}
{Lyne} A.~G.,  {Manchester} R.~N.,   {D'Amico} N.,  1996, ApJ, 460, L41

\bibitem[\protect\citeauthoryear{{Lyne} et~al.,}{{Lyne} et~al.}{2004}]{lbk+04}
{Lyne} A.~G.,  et~al., 2004, Science, 303, 1153

\bibitem[\protect\citeauthoryear{{Lyne}, {Hobbs}, {Kramer}, {Stairs}  \&
  {Stappers}}{{Lyne} et~al.}{2010}]{lhk+10}
{Lyne} A.,  {Hobbs} G.,  {Kramer} M.,  {Stairs} I.,   {Stappers} B.,  2010,
  Science, 329, 408

\bibitem[\protect\citeauthoryear{{Lyne}, {Graham-Smith}, {Weltevrede},
  {Jordan}, {Stappers}, {Bassa}  \& {Kramer}}{{Lyne} et~al.}{2013}]{lgw+13}
{Lyne} A.,  {Graham-Smith} F.,  {Weltevrede} P.,  {Jordan} C.,  {Stappers} B.,
  {Bassa} C.,   {Kramer} M.,  2013, Science, 342, 598

\bibitem[\protect\citeauthoryear{{Manchester} et~al.,}{{Manchester}
  et~al.}{1996}]{mld+96}
{Manchester} R.~N.,  et~al., 1996, MNRAS, 279, 1235

\bibitem[\protect\citeauthoryear{{McSweeney}, {Bhat}, {Wright}, {Tremblay}  \&
  {Kudale}}{{McSweeney} et~al.}{2019}]{sbw+19}
{McSweeney} S.~J.,  {Bhat} N.~D.~R.,  {Wright} G.,  {Tremblay} S.~E.,
  {Kudale} S.,  2019, ApJ, 883, 28

\bibitem[\protect\citeauthoryear{{Ng} \& {Romani}}{{Ng} \&
  {Romani}}{2004}]{Ng2004}
{Ng} C.~Y.,  {Romani} R.~W.,  2004, ApJ, 601, 479

\bibitem[\protect\citeauthoryear{{Noutsos}, {Johnston}, {Kramer}  \&
  {Karastergiou}}{{Noutsos} et~al.}{2008}]{njkk08}
{Noutsos} A.,  {Johnston} S.,  {Kramer} M.,   {Karastergiou} A.,  2008, MNRAS,
  386, 1881

\bibitem[\protect\citeauthoryear{{Olszanski}, {Mitra}  \& {Rankin}}{{Olszanski}
  et~al.}{2019}]{omr19}
{Olszanski} T. E.~E.,  {Mitra} D.,   {Rankin} J.~M.,  2019, MNRAS, 489, 1543

\bibitem[\protect\citeauthoryear{{Oswald}, {Karastergiou}  \&
  {Johnston}}{{Oswald} et~al.}{2019}]{okj19}
{Oswald} L.,  {Karastergiou} A.,   {Johnston} S.,  2019, MNRAS, 489, 310

\bibitem[\protect\citeauthoryear{{{\"O}zel} \& {Freire}}{{{\"O}zel} \&
  {Freire}}{2016}]{of16}
{{\"O}zel} F.,  {Freire} P.,  2016, ARA\&A, 54, 401

\bibitem[\protect\citeauthoryear{{Palfreyman}, {Dickey}, {Hotan}, {Ellingsen}
  \& {van Straten}}{{Palfreyman} et~al.}{2018}]{pdh+18}
{Palfreyman} J.,  {Dickey} J.~M.,  {Hotan} A.,  {Ellingsen} S.,   {van Straten}
  W.,  2018, Nature, 556, 219

\bibitem[\protect\citeauthoryear{{Parthasarathy} et~al.,}{{Parthasarathy}
  et~al.}{2019}]{psj+19}
{Parthasarathy} A.,  et~al., 2019, MNRAS, 489, 3810

\bibitem[\protect\citeauthoryear{{Pennucci}}{{Pennucci}}{2019}]{pen19}
{Pennucci} T.~T.,  2019, ApJ, 871, 34

\bibitem[\protect\citeauthoryear{{Perera}, {Stappers}, {Weltevrede}, {Lyne}  \&
  {Rankin}}{{Perera} et~al.}{2016}]{psw+16}
{Perera} B.~B.~P.,  {Stappers} B.~W.,  {Weltevrede} P.,  {Lyne} A.~G.,
  {Rankin} J.~M.,  2016, MNRAS, 455, 1071

\bibitem[\protect\citeauthoryear{{Petroff}, {Keith}, {Johnston}, {van Straten}
  \& {Shannon}}{{Petroff} et~al.}{2013}]{pkj+13}
{Petroff} E.,  {Keith} M.~J.,  {Johnston} S.,  {van Straten} W.,   {Shannon}
  R.~M.,  2013, MNRAS, 435, 1610

\bibitem[\protect\citeauthoryear{{Philippov}, {Spitkovsky}  \&
  {Cerutti}}{{Philippov} et~al.}{2015}]{psc15}
{Philippov} A.~A.,  {Spitkovsky} A.,   {Cerutti} B.,  2015, ApJ, 801, L19

\bibitem[\protect\citeauthoryear{{Pierbattista}, {Harding}, {Grenier},
  {Johnson}, {Caraveo}, {Kerr}  \& {Gonthier}}{{Pierbattista}
  et~al.}{2015}]{Pierbattista2015}
{Pierbattista} M.,  {Harding} A.~K.,  {Grenier} I.~A.,  {Johnson} T.~J.,
  {Caraveo} P.~A.,  {Kerr} M.,   {Gonthier} P.~L.,  2015, A\&A, 575, A3

\bibitem[\protect\citeauthoryear{{Radhakrishnan} \& {Cooke}}{{Radhakrishnan} \&
  {Cooke}}{1969}]{rc69}
{Radhakrishnan} V.,  {Cooke} D.~J.,  1969, Ap. Lett., 3, 225

\bibitem[\protect\citeauthoryear{{Radhakrishnan} \&
  {Manchester}}{{Radhakrishnan} \& {Manchester}}{1969}]{rm69}
{Radhakrishnan} V.,  {Manchester} R.~N.,  1969, Nature, 222, 228

\bibitem[\protect\citeauthoryear{{Rankin}}{{Rankin}}{1986}]{ran86}
{Rankin} J.~M.,  1986, ApJ, 301, 901

\bibitem[\protect\citeauthoryear{{Rankin}}{{Rankin}}{1993}]{ran93}
{Rankin} J.~M.,  1993, ApJ, 405, 285

\bibitem[\protect\citeauthoryear{{Rankin}, {Rodriguez}  \& {Wright}}{{Rankin}
  et~al.}{2006}]{rrw06}
{Rankin} J.~M.,  {Rodriguez} C.,   {Wright} G.~A.~E.,  2006, MNRAS, 370, 673

\bibitem[\protect\citeauthoryear{{Rigoselli}, {Mereghetti}, {Turolla},
  {Taverna}, {Suleimanov}  \& {Potekhin}}{{Rigoselli}
  et~al.}{2019}]{Rigoselli2019}
{Rigoselli} M.,  {Mereghetti} S.,  {Turolla} R.,  {Taverna} R.,  {Suleimanov}
  V.,   {Potekhin} A.~Y.,  2019, ApJ, 872, 15

\bibitem[\protect\citeauthoryear{{Rookyard}, {Weltevrede}  \&
  {Johnston}}{{Rookyard} et~al.}{2015}]{rwj15a}
{Rookyard} S.~C.,  {Weltevrede} P.,   {Johnston} S.,  2015, MNRAS, 446, 3367

\bibitem[\protect\citeauthoryear{{Ruderman} \& {Sutherland}}{{Ruderman} \&
  {Sutherland}}{1975}]{rs75}
{Ruderman} M.~A.,  {Sutherland} P.~G.,  1975, ApJ, 196, 51

\bibitem[\protect\citeauthoryear{{Shannon} \& {Cordes}}{{Shannon} \&
  {Cordes}}{2010}]{sc10}
{Shannon} R.~M.,  {Cordes} J.~M.,  2010, ApJ, 725, 1607

\bibitem[\protect\citeauthoryear{{Shannon} et~al.,}{{Shannon}
  et~al.}{2015}]{srl+15}
{Shannon} R.~M.,  et~al., 2015, Science, 349, 1522

\bibitem[\protect\citeauthoryear{{Sobey} et~al.,}{{Sobey}
  et~al.}{2019}]{sbg+19}
{Sobey} C.,  et~al., 2019, MNRAS, 484, 3646

\bibitem[\protect\citeauthoryear{{Sotomayor-Beltran}
  et~al.,}{{Sotomayor-Beltran} et~al.}{2013}]{ssh+13}
{Sotomayor-Beltran} C.,  et~al., 2013, A\&A, 552, A58

\bibitem[\protect\citeauthoryear{{Stairs} et~al.,}{{Stairs}
  et~al.}{2019}]{slk+19}
{Stairs} I.~H.,  et~al., 2019, MNRAS, 485, 3230

\bibitem[\protect\citeauthoryear{{Szary} \& {van Leeuwen}}{{Szary} \& {van
  Leeuwen}}{2017}]{svl17}
{Szary} A.,  {van Leeuwen} J.,  2017, ApJ, 845, 95

\bibitem[\protect\citeauthoryear{{Thornton} et~al.,}{{Thornton}
  et~al.}{2013}]{tsb+13}
{Thornton} D.,  et~al., 2013, Science, 341, 53

\bibitem[\protect\citeauthoryear{{Wahl}, {Orfeo}, {Rankin}  \&
  {Weisberg}}{{Wahl} et~al.}{2016}]{worw16}
{Wahl} H.~M.,  {Orfeo} D.~J.,  {Rankin} J.~M.,   {Weisberg} J.~M.,  2016,
  MNRAS, 461, 3740

\bibitem[\protect\citeauthoryear{{Walker}, {Koopmans}, {Stinebring}  \& {van
  Straten}}{{Walker} et~al.}{2008}]{wks+08}
{Walker} M.~A.,  {Koopmans} L.~V.~E.,  {Stinebring} D.~R.,   {van Straten} W.,
  2008, MNRAS, 388, 1214

\bibitem[\protect\citeauthoryear{{Walker}, {Demorest}  \& {van
  Straten}}{{Walker} et~al.}{2013}]{wdv13}
{Walker} M.~A.,  {Demorest} P.~B.,   {van Straten} W.,  2013, ApJ, 779, 99

\bibitem[\protect\citeauthoryear{{Weltevrede}, {Edwards}  \&
  {Stappers}}{{Weltevrede} et~al.}{2006}]{wes06}
{Weltevrede} P.,  {Edwards} R.~T.,   {Stappers} B.~W.,  2006, \mn@doi [A\&A]
  {10.1051/0004-6361:20053088}, \href
  {https://ui.adsabs.harvard.edu/abs/2006A&A...445..243W} {445, 243}

\bibitem[\protect\citeauthoryear{{Weltevrede} et~al.,}{{Weltevrede}
  et~al.}{2010}]{wjm+10}
{Weltevrede} P.,  et~al., 2010, PASA, 27, 64

\bibitem[\protect\citeauthoryear{{Wright} \& {Weltevrede}}{{Wright} \&
  {Weltevrede}}{2017}]{ww17}
{Wright} G.,  {Weltevrede} P.,  2017, MNRAS, 464, 2597

\bibitem[\protect\citeauthoryear{{Yao}, {Manchester}  \& {Wang}}{{Yao}
  et~al.}{2017}]{ymw17}
{Yao} J.~M.,  {Manchester} R.~N.,   {Wang} N.,  2017, ApJ, 835, 29

\bibitem[\protect\citeauthoryear{{van Straten} \& {Bailes}}{{van Straten} \&
  {Bailes}}{2011}]{vb11}
{van Straten} W.,  {Bailes} M.,  2011, PASA, 28, 1

\makeatother
\end{thebibliography}

%%%%%%%%%%%%%%%%%%%%%%%%%%%%%%%%%%%%%%%%%%%%%%%%%%

% Don't change these lines
\bsp	% typesetting comment
\label{lastpage}
\end{document}